\documentclass[aps,prl,amsmath,showpacs,preprintnumbers,
               floatfix,twocolumn]{revtex4}
\usepackage{amsmath}
\usepackage{graphicx}
\usepackage{latexsym}

\newcommand{\beqa}{\begin{eqnarray}}
\newcommand{\eeqa}{\end{eqnarray}}
\newcommand{\beqn}{\begin{equation}}
\newcommand{\eeqn}{\end{equation}}

\def\spose#1{\hbox to 0pt{#1\hss}}
\def\ltapprox{\mathrel{\spose{\lower 3pt\hbox{$\mathchar"218$}}
 \raise 2.0pt\hbox{$\mathchar"13C$}}}
\def\gtapprox{\mathrel{\spose{\lower 3pt\hbox{$\mathchar"218$}}
 \raise 2.0pt\hbox{$\mathchar"13E$}}}

\begin{document}

\author{Wanderson G. Wanzeller}
\affiliation{Instituto de F\'\i sica Te\'orica, Universidade
Estadual Paulista, Rua Pamplona 145, 01405-900 S\~ao Paulo, SP,
Brazil }
\title{Short-time Dynamics of Percolation Observables}
\author{Tereza Mendes}
\affiliation{Instituto de F\'\i sica de S\~ao Carlos,
Universidade de S\~ao Paulo \\
C.P.\ 369, 13560-970 S\~ao Carlos, SP, Brazil}
\author{Gast\~ao Krein}
\affiliation{Instituto de F\'\i sica Te\'orica,
Universidade Estadual Paulista,\\
Rua Pamplona 145, 01405-900 S\~ao Paulo, SP, Brazil
}

\begin{abstract}
\noindent
We consider the critical short-time evolution of magnetic and
droplet-percolation order parameters for the Ising model
in two and three dimensions, through Monte-Carlo simulations with the
(local) heat-bath method.
We find qualitatively different dynamic behaviors for the two types of
order parameters. More precisely, we find that the percolation order parameter
does not have a power-law behavior as encountered for the magnetization, but
develops a scale (related to the relaxation time to equilibrium)
in the Monte-Carlo time.
We argue that this difference is due to the difficulty in forming
large clusters at the early stages of the evolution.
Our results show that, although the descriptions
in terms of magnetic and percolation order parameters may be equivalent
in the equilibrium regime, greater care must be taken to interprete percolation
observables at short times.
In particular, this concerns the attempts to describe the dynamics of the
deconfinement phase transition in QCD using cluster observables.

\end{abstract}

\pacs{11.10.Cd, 11.15.Ha, 12.38.Aw, 14.70.Dj}

\maketitle

\section{Introduction}
It is often useful to map a physical phase transition into the
geometric problem of percolation, e.g.\ in order to gain a better
understanding of how the transition is induced in the system (see
e.g.\ \cite{Wenzel:2005nd}). This mapping is obtained through a
suitable definition of {\em cluster} in terms of the system's
variables and parameters. In the case of the Ising model, such
physical (droplet) clusters were introduced by Coniglio and Klein
(based on the prescription by Kasteleyn and Fortuin)~\cite{CK} and
the mapping is well understood \cite{stauffer}. This holds also for
the so-called $n$-vector models~\cite{Blanchard:2000fj}. The correct
cluster definition is of special interest in the description of more
complex phase transitions, such as the quark-deconfinement
transition in finite-temperature QCD, for which the physical order
parameter is not yet established. In this case there is still no
satisfactory definition for a ``droplet'' cluster, even in the
simpler pure-gauge case \cite{satz}.

When equilibrium properties are investigated, the descriptions of
the spin-model phase transition in terms of the magnetic order
parameter or the percolation order parameter are indeed equivalent
and one finds the same critical exponents
(see, for example, \cite{BJP} and references therein).
The same may or may not be true for the dynamical evolution of the
different types of order parameters, although in principle
one might expect to find equivalence for dynamic quantities as well,
in particular for the behavior at short times.
We recall that the short-time behavior of the magnetic order parameter
$M$ is described by a scale-free expression, in terms of a power law
(see for instance \cite{zheng0,zheng,zheng_rev}).

The study of the dynamic critical behavior of a percolation order
parameter with respect to the (Glauber) Monte-Carlo evolution might
be of relevance for understanding
non-equilibrium effects in hot QCD, such as the effects due to heating
and cooling of matter produced in heavy-ion collisions.
Indeed, the possible connection between the deconfinement
transition in QCD and the percolation phenomenon \cite{baym} has
received renewed attention in recent years \cite{satz} and the dynamics
of cluster observables has been investigated using hysteresis methods
\cite{hyst}.
Note that the QCD phase transition is predicted to fall into the Ising
universality class in the pure two-color [$SU(2)$] case and that
the chiral phase transition in the two-flavor full-QCD case is expected
to be in the universality class of the (continuous-spin) 4-vector model
\cite{sigma,o4,o4o2}. This motivates the connection between the percolation
transitions for spin models and the (dynamic) behavior at the QCD phase
transition. We thus consider
here the short-time dynamics of the two- and three-dimensional
Ising model and focus on
the dynamic critical behavior of the percolation order parameter.
(Preliminary results of our study were presented in \cite{prelim,prelim2}.)
Note that the dynamic behavior of cluster observables has been considered
for the droplet clusters in Ising and Potts models in various
studies (see e.g.\ \cite{stauffer2} and \cite{yasar}), but mostly
for the so-called cluster numbers $n_s$ --- which denote the average
number of clusters (per lattice site) containing $s$ sites each --- 
and not for the percolation order parameter.

Here we investigate the dynamic critical behavior of the
(zero-field) Ising model
\begin{equation}
{\cal H} \;=\; -J \sum_{<i,j>} S_i\,S_j\,,
\end{equation}
where $J$ is positive and each $S_i = \pm 1$,
for short Monte-Carlo times $t$, using the (local) heat-bath algorithm.
We measure the magnetization $M\,=\,(1/V)\,\sum_{i}S_i$
and we consider two different definitions for the percolation order parameter.
Indeed, given a definition for a cluster on the lattice, one may consider
as the order parameter in percolation theory the
stress of the percolating cluster, defined for each configuration by
the relative volume of the infinite cluster.
(In particular, in the droplet picture for the Ising model, this would
give zero for temperatures above the critical one.)
On a finite lattice, one might consider the volume of the infinite cluster
to be the one of the spanning cluster, or zero when there is no
percolating cluster (i.e.\ above the critical temperature).
Note that a spanning cluster is a set of spins
connected from the first to the last row of the lattice
in at least one of its space directions.
Alternatively, one can also consider the relative volume of the
largest cluster (see e.g.\ Chapter 3 in \cite{stauffer}).
We denote these two definitions for the percolation order parameter
respectively by $\Omega$ (in the case of
the spanning cluster) and $\Omega'$ (in the case of the largest cluster).

In this work we
compare the behaviors of the two types of quantities just described
(i.e., percolation or magnetic order parameters) as functions of $t$.
As shown below, we find that
whereas the magnetic order parameter displays a power-law increase with $t$,
the data for the percolation order parameters $\Omega$ and $\Omega'$
are well fitted by a diffusion and by a growth and nucleation process
respectively. In both cases the time scales --- respectively called $\tau$
and $\tau'$ --- are
related to the relaxation time to equilibrium. Indeed, we show that
$\tau$ and $\tau'$ diverge as $L^z$ when the lattice side $L$ tends to
infinity, where $z\approx 2$ is the dynamic critical exponent of the
heat-bath algorithm \cite{drugo}.

\section{Short-time (Monte Carlo) dynamics}
Using renormalization-group theory, it can be shown \cite{janssen,zheng_rev}
that the early time evolution of an order parameter (e.g.\ the magnetization
$M$) already displays universal critical behavior, given by
\begin{eqnarray}
M(t,\epsilon,m_0) = b^{-\beta/\nu}{\cal M}(t b^{-z},
\epsilon b^{1/\nu},m_0 b^{x_0}) \,,
\end{eqnarray}
where $m_0$ is the initial magnetization, $\epsilon \equiv (T-T_c)/T_c \,$,
${\cal M}$ is a universal function and $b$ is a scale factor, which can be
taken equal to $t^{1/z}$. We thus expect for $T=T_c$ and small $m_0$ a power-law
behavior at early times
\begin{equation}
M(t)_{\epsilon \rightarrow 0} \sim m_0 t^\theta\,,
\end{equation}
with $\theta=(x_0-\beta/\nu)/z\,$.
In principle, we would assume that the two percolation order parameters
{$\Omega$} or $\Omega'$ defined above should have a similar behavior.

The heat-bath dynamics
consists in choosing the two possible
directions of each Ising spin according to the exact conditional probability
given by its nearest neighbors.
Each spin $S_i$ is chosen ``up'' or ``down'' respectively with
probability ${p_i}$ or ${1-p_i}$, where
\begin{eqnarray}
p_i = \frac{1}{1+\exp(-2\beta \sum_{j} S_j)}
\end{eqnarray}
and the sum is over nearest neighbors of $S_i$.
After a certain number of iterations the spin configuration obeys the
Boltzmann distribution. In the heat-bath method, since the updates are
local, this transient time becomes considerably large at criticality.

The Fortuin-Kasteleyn clusters are obtained from
the Ising-model Hamiltonian by writing the partition function as
\begin{eqnarray}
{\cal Z}=\sum_{\{S\}}\sum_{\{n\}}\left\{
\prod_{\langle i,j\rangle}^{n_{ij}=1} p_{ij}\delta_{S_iS_j}\right \}
\left\{ \prod_{\langle i,j\rangle}^{n_{ij}=0}(1-p_{ij})   \right \}\,,
\end{eqnarray}
where $p_{ij}=1-\exp(-2J\beta)$ is the probability of having a
link between two nearest-neighbor sites of equal spin value. This
link is represented by ${n_{ij}}$ and determines the
clusters that will be associated with percolation at the critical
temperature \cite{sokal}.
Note that the above defined clusters are used in the Swendsen-Wang
algorithm \cite{SW} to perform
{\em global} moves in which the spins in a cluster are flipped together.
Here we only use these clusters to calculate percolation
observables, whereas the dynamics is given by local heat-bath updates,
as described above.

\section{Numerical results}
In order to study the short-time dynamics we simulate at {$T=T_c$}
and force the system to have an initial magnetization {$m_0$}.
We let the system evolve in time and look for power-law behavior of the
order parameters $M$, $\Omega$ and $\Omega'$ as functions of the
(Monte-Carlo) time.
Each temporal sequence is generated from a different random seed,
i.e.\ each sequence has a different initial spin configuration.
The time history is then obtained from an average over all the
generated sequences.

We have studied the two- and three-dimensional cases, performing
Monte Carlo simulations respectively with 50,000 and 40,000 seeds,
for several initial magnetizations $m_0$
and several lattice volumes, using the heat-bath algorithm.
We consider periodic boundary conditions.
Note that in order to compare the percolation order parameters
$\Omega$ and $\Omega'$ to $M$ we consider the volumes of clusters
of ``up'' spins with positive sign and volumes of clusters of ``down''
spins with negative sign when taking the average over the seeds.

\begin{figure}[ht]
\protect\vspace*{-0cm}
\includegraphics[height=0.7\hsize,angle=-0]{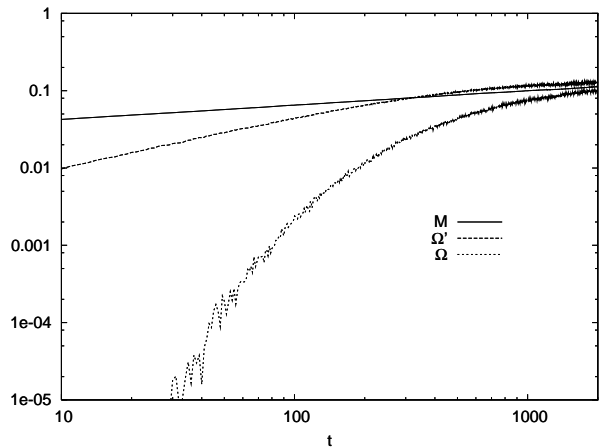}
\caption{ Plot of the early time evolution of the magnetic ($M$) and
percolation ($\Omega$, $\Omega'$) order parameters for the
two-dimensional case. Data are shown for $m_0 = 0.02$ and $L=200$.
Note the logarithmic scale on both axes.} \label{2d}
\end{figure}

\begin{figure}[ht]
\protect\vspace*{-0cm}
\protect\hspace*{-0.5cm}
\includegraphics[height=0.7\hsize,angle=0]{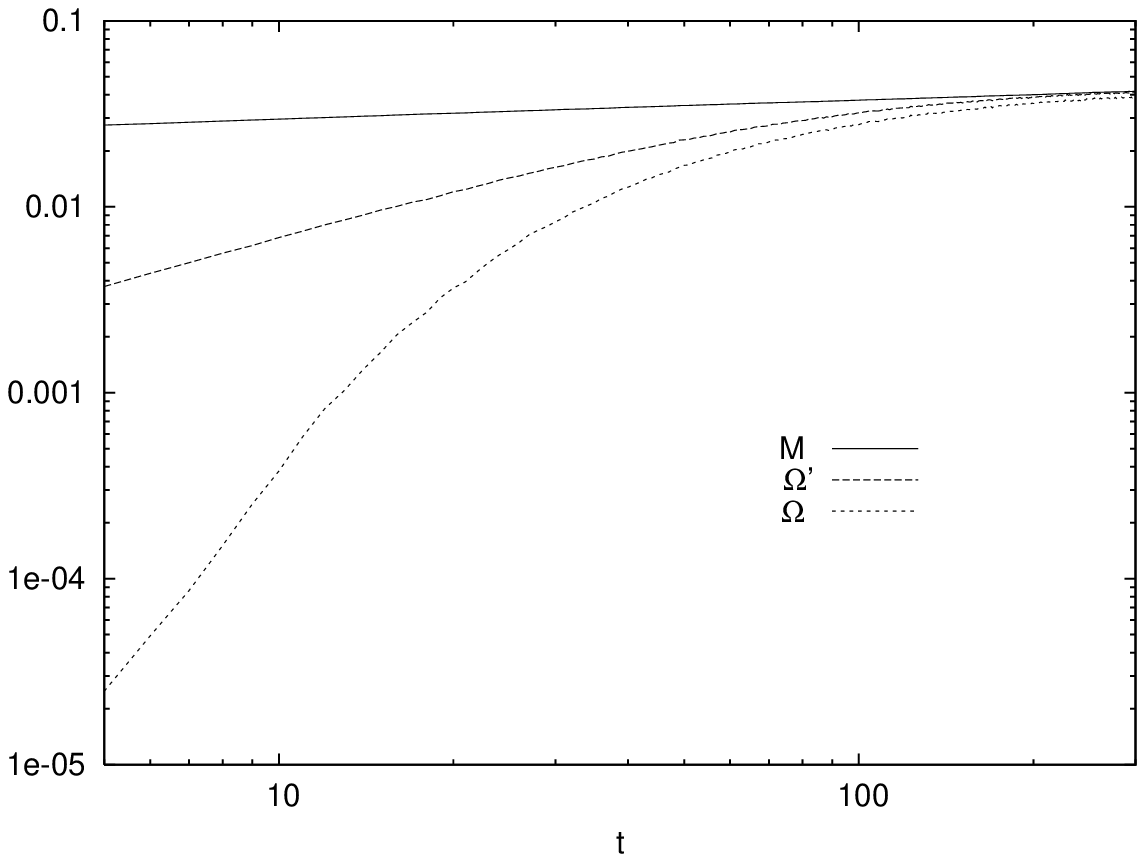}
\caption{
Plot of the early time evolution of the magnetic ($M$)
and percolation ($\Omega$, $\Omega'$) order parameters for
the three-dimensional case. Data are shown for $m_0 = 0.02$ and $L=64$.
Note the logarithmic scale on both axes.}
\label{3d}
\end{figure}

We obtain that a power-law fit works very well for $M$, yielding the
literature value \cite{zhang} for the exponent $\theta$. However, as
can be seen in Figs.\ \ref{2d} and \ref{3d}, the percolation order
parameters $\Omega$ and $\Omega'$ do not show a power-law behavior,
being consistent with exponential behaviors in terms of $t/\tau$,
thus having $\tau$ as a time scale. As verified below, the
exponential behaviors are different for $\Omega$ and $\Omega'$, but
$\tau$ is in both cases directly related to the relaxation time to
equilibrium. We thus find the surprising result that although the
equilibrium behaviors of $M$, $\Omega$ and $\Omega'$ are equivalent,
the different types of order parameters show qualitatively different
dynamic critical behavior.

More precisely, the behavior of $\Omega$ is given in the two- and
three-dimensional cases by
\begin{equation}
\Omega(t) \;=\; A\,\exp(-\tau/t)\,, \label{exp}
\end{equation}
as can be seen from Table \ref{omega}.
On the other hand, the behavior of $\Omega'$ is given by
\begin{equation}
\Omega'(t) \;=\; B\,\left\{1\,-\,\exp\left[-(t/\tau')^\eta\right]\right\}\, ,
\label{exp2}
\end{equation}
as shown in Table \ref{omega'}.

\begin{table} 
\caption{Fits of $\Omega(t)$ to the form $A\,\exp(-\tau/t)$. Data for the
case $m_0 = 0.02$. The fit intervals are chosen as shown in the
second column, whereas the errors in the fit parameters are adjusted to account
for the fluctuations in these parameters for slightly different fit intervals.}
\begin{ruledtabular}
\begin{tabular}{c|c|c|c|c}
Volume & $\Delta t$ & $\qquad$A$\qquad$ & $\qquad\tau\qquad$ & $\chi^2/d.o.f.$ \\
\hline
$100^2$ & 18--83 & 0.061(1) & 85.5(9) & 0.97 \\
$125^2$ & 28--110 & 0.062(1) & 134(1) & 0.98 \\
$150^2$ & 30-150 & 0.063(1) &  192(2) & 0.91 \\
$200^2$ & 41-240 & 0.061(1) & 334(3)  & 0.69 \\  \hline
$24^3$  & 3--100 & 0.0366(3)  & 5.9(3)  & 0.57 \\
$32^3$  & 10--120 & 0.0397(5) & 11.7(6) & 0.42 \\
$48^3$  & 15--120 & 0.0446(4) & 28.4(4) & 0.58 \\
$64^3$  & 20--130 & 0.0467(3) & 51.6(3) & 0.68 \\
\end{tabular}
\end{ruledtabular}
\label{omega}
\end{table}

\begin{table}[htb]
\caption{Fits of $\Omega'(t)$ to the form $B\,(1\,-\,
\exp[-(t/\tau')^\eta])$. Data for the case $m_0 = 0.02$. The fit
intervals are chosen as shown in the second column, whereas the
errors in the fit parameters are adjusted to account for the
fluctuations in these parameters for slightly different fit
intervals.}
\begin{ruledtabular}
\begin{tabular}{c|c|c|c|c|c}
Volume & $\Delta t$ & $\qquad$B$\qquad$ & $\quad\tau'\quad$ &
$\quad\eta\quad$ & $\chi^2/d.o.f.$ \\
$100^2$ & 7--100 & 0.104(5) & 82(1)   & 0.71(1) & 0.35 \\
$125^2$ & 8--100 & 0.126(5) & 168(14) & 0.70(1) & 0.23 \\
$150^2$ & 7--100 & 0.115(4) & 179(13) & 0.72(1) & 0.17 \\
$200^2$ & 10--100 & 0.20(1) & 380(32) & 0.73(1) & 0.17 \\  \hline
$24^3$ & 3--100 & 0.0349(1) & 8.0(1)  & 0.82(1) & 0.42 \\
$32^3$ & 5--120 & 0.0372(2) & 15.4(3) & 0.83(2) & 0.50 \\
$48^3$ & 10--130 & 0.0394(3) & 34.4(4) & 0.90(1) &  0.48 \\
$64^3$ & 15--150 & 0.0402(5) & 60.5(4)  & 0.93(1) & 0.56 \\
\end{tabular}
\end{ruledtabular}
\label{omega'}
\end{table}

\begin{figure}[ht]
\protect\vspace*{-0cm}
\protect\hspace*{-0.5cm}
\includegraphics[height=0.7\hsize,angle=-0]{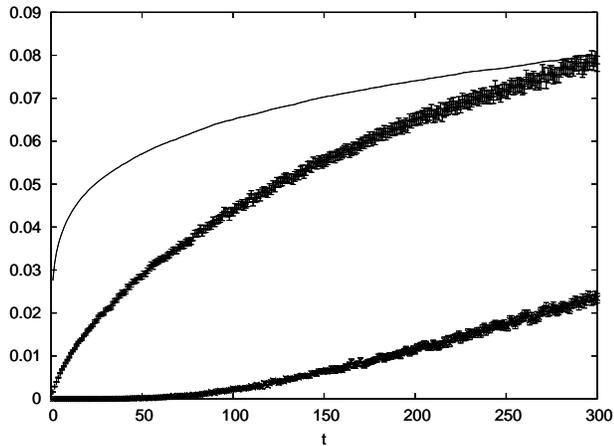}
\caption{
Data for the $2d$ case
and fits of $\Omega$ (lowest curve) and $\Omega'$ (middle curve)
according to the forms in Eqs.\ (\ref{exp}) and (\ref{exp2}) respectively.
The magnetization (top curve) is also shown, for comparison.
Data and fits are shown for $m_0 = 0.02$ and $L=200$.
Error bars correspond to one standard deviation.}
\label{2dfits}
\end{figure}

\begin{figure}[ht]
\protect\vspace*{-0cm}
\protect\hspace*{-0.5cm}
\includegraphics[height=0.7\hsize,angle=0]{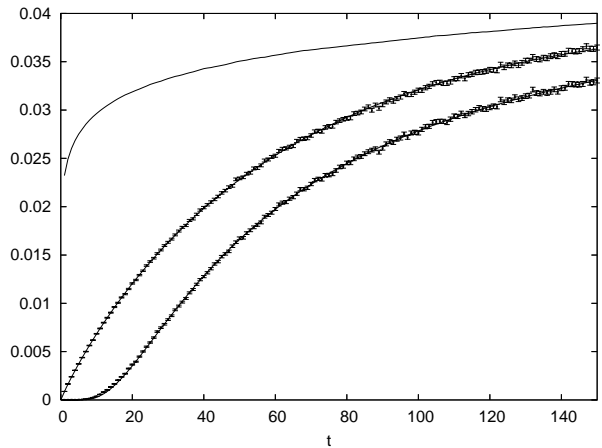}
\caption{
Data for the $3d$ case
and fits of $\Omega$ (lowest curve) and $\Omega'$ (middle curve)
according to the forms in Eqs.\ (\ref{exp}) and (\ref{exp2}) respectively.
The magnetization (top curve) is also shown, for comparison.
Data and fits are shown for $m_0 = 0.02$ and $L=64$.
Error bars correspond to one standard deviation.}
\label{3dfits}
\end{figure}

Fits of the data to the above forms are shown (for the largest
lattices considered) in Figs.\ \ref{2dfits} and \ref{3dfits}
respectively for the $2d$ and $3d$ cases. We see that the departure
of the percolation order parameters from the power-law behavior
of the magnetization remains even long after the so-called
microscopic time \cite{zheng_rev}. In other words, neither $\Omega$
nor $\Omega'$ can be fitted to a power law at short times. In fact,
the forms proposed for $\Omega$ and $\Omega'$ in Eqs.\ (\ref{exp}) and
(\ref{exp2}) are well
fitted by the data in two and three dimensions, for all volumes
considered. (We note however that
the two-dimensional data for $\Omega$ were fitted in \cite{prelim2}
considering also a second
exponential term, adding a third parameter to this fit.)
The above data for $\tau$ and $\tau'$ can be fitted to the
form $L^z$, in the two- and three-dimensional cases separately.
In all four cases we find $z\approx 2$.
The two time scales can therefore
be associated with the relaxation time to equilibrium.

Note that the behavior of $\Omega$ (in two and three dimensions) is
a factor $\sqrt{t}$ times the solution of a one-dimensional
diffusion equation at a fixed position in space \cite{doremus}. In
fact, the expression for the concentration of diffusion material due
to a change in chemical potential along a direction $x$ at a
position $x_0$ and time $t$ is given by
$(C/\sqrt{t})\,\exp(-x_0^2/4Dt)\,,$ where $D$ is the diffusion
coefficient. Comparing this to the form for $\Omega$ in Eq.\ (\ref{exp}), 
we see that if we multiply the percolation order parameter
in this case by $\sqrt{t}$ we get the diffusive form, identifying
$x_0^2/4D$ with $\tau$. Of course, the factor $\sqrt{t}$ corresponds
to the diffusion length for a random walk at time $t$. Moreover, the
fixed position $x_0$ is proportional to $\sqrt{\tau}$, i.e.\ roughly the
length of the lattice (since $\tau\approx L^z$ as mentioned above). 
This connects the occurrence and strength of
percolation at time $t$ to the probability that a random walk will
reach a length $L$ after $t$ steps.

The behavior of $\Omega'$ corresponds to the volume fraction of
particles in a process of nucleation and growth \cite{doremus}. 
This behavior is observed in the dynamics of weak first-order
phase transitions as seen e.g.\ in Ref.~\cite{gleiser}. In
that reference, the critical dynamics of a scalar field
quenched to a metastable state was investigated with a model-A
Langevin equation. In this context, the result for $\Omega'$ is
particularly illuminating in the sense that for the same underlying
dynamics (model A, in this case of the heath-bath dynamics), the
short-time behavior clearly depends on the specific observable and
its initial configuration. It would be also interesting to check if
the exponent $\eta$ goes to 1 in the limit of very large $L$ in
three dimensions (as seems to be suggested by the data in
Table~\ref{omega'}) or to relate the exponents in two and three
dimensions to the fractal dimensions associated with the percolating
clusters in the two cases.

Regarding the so-called percolation cumulant or percolation
probability --- taken as 1 if there is percolation and 0 if there is
not --- we obtain that this quantity does not show a power-law behavior
in time, contrary to what is observed for the Binder cumulant
\cite{zheng_rev}. We find that the percolation cumulant is also described 
by an exponential $\exp(-\tau/t)$, with the prefactor $\sqrt{t}$ in two 
dimensions and $t^{-0.34}$ in three dimensions.


\section{Conclusions}
We have investigated numerically the critical heat-bath dynamics for
magnetic and percolation order parameters in the Ising model at
short Monte Carlo times, starting from a small magnetization $m_0$.
From our results we see that although the equilibrium behaviors of
the magnetization $M$ and of the percolation order parameters
$\Omega$ and $\Omega'$ are equivalent, the two types of order
parameters show qualitatively different dynamic critical behavior at
short times. Indeed, whereas the magnetic order parameter $M$ shows a
power-law behavior with the exponent $\theta$, one finds that
$\Omega$ and $\Omega'$ have a time scale, given respectively by
$\tau$ in Eq.~(\ref{exp}) and $\tau'$ in Eq.~(\ref{exp2}). This time
scale seems to be related to the relaxation time of the algorithm
used for thermalization.

The short-time behaviors of $\Omega$ and $\Omega'$ are well described
respectively by diffusion and by growth and nucleation processes,
probably related to the difficulty in forming a percolating cluster at the
early stages of the simulation.


\section*{ACKNOWLEDGMENTS}

We thank Attilio Cucchieri for helpful discussions.
Research supported by FAPESP, CNPq and CAPES.


\end{document}